\documentclass[aps,showpacs,twocolumn,showkeys,amsmath,amssymb,floatfix,
nofootinbib,preprintnumbers]
{revtex4}
\usepackage{bm}
\usepackage{epsfig}

\begin{document}

\title{The QCD Dynamics of Tetraquark Production}

\author{Stanley J. Brodsky}
\email{sjbth@slac.stanford.edu}
\affiliation{SLAC National Accelerator Laboratory, Stanford
University, Stanford, California 94039, USA}

\author{Richard F. Lebed}
\email{richard.lebed@asu.edu}
\affiliation{Department of Physics, Arizona State University, Tempe,
Arizona 85287-1504, USA}

\date{May, 2015}

\begin{abstract}
  We use the twist dimensions of the operators underlying the
  dynamical behavior of exclusive production processes as a tool for
  determining the structure of exotic heavy-quark states such as the
  $Z_c^+(4430)$ tetraquark.  The resulting counting rules predict
  distinctive fall-offs of the cross sections in center-of-mass
  energy, thus distinguishing whether the tetraquarks are segregated
  into di-meson molecules, diquark-antidiquark pairs, or more
  democratically arranged four-quark states.  In addition, we propose
  straightforward methods of experimentally producing additional
  exotic multiquark states.
\end{abstract}

\preprint{SLAC-PUB-16259}


\maketitle


\section{Introduction} \label{sec:Intro}

Hadronic physics has reached an important milestone in the past year:
the experimental confirmation by LHCb~\cite{Aaij:2014jqa} of the {\em
tetraquark\/} ($\bar c c \bar d u$) state $Z_c^+ (4430)$ with
spin-parity $J^P = 1^+$.  Its interpretation as a true resonance is
confirmed by the observation that the phase shift $\delta$ of its
complex production amplitude increases by $\frac \pi 2$ radians as the
energy crosses the resonant mass.  The tetraquark thus joins the $\bar
q q$ meson and $qqq$ baryon as a third class of hadrons.  The
$Z_c^+(4430)$ represents just one of a growing collection of
unexpected charmonium-like states, beginning with the famous $X(3872)$
first seen by Belle in 2003~\cite{Choi:2003ue}.

The discovery of the charged ($Z_c^+$) states requires a valence quark
content of at least four quarks, $\bar c c \bar d u$, and the strength
of the observed transitions amongst the $X$, $Y$ (charmonium-like
states appearing in initial-state radiation processes $e^+ e^- \to
\gamma Y$), and $Z_c$ strongly suggest a common tetraquark nature for
all of these novel states~\cite{Yuan:2014rta}.  The $J^{PC} = 1^{++}$
$X(3872)$, for example, is almost certainly a $\bar c c \bar q q$
state, in which $\bar q q$ is a linear combination of $\bar u u$ and
$\bar d d$.

In this paper we will discuss two essential questions: (a) the color
composition of the tetraquarks in QCD, and (b) the dynamics underlying
their production.  For example, we shall argue that tetraquarks such
as the $Z_c$ can be produced near threshold in both hadronic
collisions and electroproduction.  We will also discuss the
possible existence of other novel multiquark hadronic states that
are natural extensions of tetraquark states.

The first, and still most widely known, ansatz proposed for the
structure of tetraquarks is one of di-meson molecules bound by pion
exchange or color van der Waals forces (as reviewed in, {\it e.g.},
Ref.~\cite{Brambilla:2014aaa}).  The proximity of several of these
states to the corresponding two-meson thresholds is quite remarkable:
For instance, $m_{X(3872)} - m_{D^{*0}} - m_{D^0} = -0.11 \pm
0.21$~MeV\@.  On the other hand, a number of other tetraquark
candidates lie just above such thresholds, suggesting 
some sort of potential barrier to allow bound states with {\em
  positive\/} binding energy, while others---prominently, the
$Z_c^+(4430)$---have no obvious nearby threshold with the appropriate
quantum numbers, in this case $J^P = 1^+$.

The prompt production cross section of the $X(3872)$ at $e^+ e^-$
colliders is substantial, indicating that the $X(3872)$ can be created
with high relative momentum between its components; however, since the
binding energy between its meson components is very small in the
di-meson molecular picture, this empirical observation creates great
difficulty for this
picture~\cite{Bignamini:2009sk,Esposito:2013ada,Guerrieri:2014gfa},
even when substantial final-state interactions between the
mesons~\cite{Artoisenet:2009wk,Artoisenet:2010uu} are taken into
account.

An alternative to the molecular picture, {\it
hadro-charm\-onium}~\cite{Voloshin:2007dx}, assumes that a compact
charmonium state is located at the center of a light-quark cloud; in
this case, one must question why such states would be quasi-stable,
and to what degree they would be obscured through mixing with
conventional charmonium.

In this work, we shall argue that tetraquarks are primarily
diquark-antidiquark ($\delta$-$\bar \delta$) bound-states.  Thus, the
$Z_c^+ (4430)$ can be considered as a $[ \bar c \bar d ]_{3C} [c
u]_{\bar 3C}$ bound state of a color-(anti)triplet charmed diquark
$\delta$ and an anti-diquark $\bar \delta$.  The same diquark clusters
appear in the valence Fock state structure of baryons; {\it e.g.}, the
$\Lambda_c(cud)$ can be considered as a color-singlet $[c u]_{\bar 3C}
d_{ 3C} + [c d]_{\bar 3C} u_{ 3C}$ composite.  Inasmuch as the
diquarks can be considered pointlike color sources, the confining
potential that confines the color-triplet $\delta$ and $\bar \delta$
into color-singlet tetraquarks is thus identical to the confinement
potential underlying $q \bar q$ mesons and $q \bar \delta$ baryons.
Thus, three strong binding interactions are present in the
diquark-antidiquark picture: the interactions creating the $\delta$
and $\bar \delta$, and the $\delta$-$\bar \delta$ binding.  This is
also the natural interpretation obtained from AdS/QCD and
superconformal algebra, an approach which accounts remarkably well for
the observed Regge spectroscopy of mesons and baryons in terms of
bound states of color-triplet and color-anti-triplet
constituents~\cite{Dosch:2015nwa,deTeramond:2014asa}.

The diquark-antidiquark picture was first proposed for charmonium-like
tetraquarks in Ref.~\cite{Maiani:2004vq}.  Since the diquarks are
color triplets, this description leads to new insights into confined
color dynamics. However, unless the model is constrained, it also
predicts many more tetraquark states than are seen experimentally.
Nevertheless, recent work~\cite{Maiani:2014aja} shows that if the
spin-spin couplings within each diquark dominate, one can explain many
empirical features of the observed tetraquark spectroscopy.  The
significance of novel color correlations in exotics such as
tetraquarks is discussed in Ref.~\cite{Karliner:2006hf}.

An important question is why the component quarks in a $\delta$-$\bar
\delta$ bound state do not immediately reorganize themselves (either
dynamically, or simply using group-theory identities) into
color-singlet $\bar q q$ pairs, thus recreating the molecular picture.
To address this objection, we note a well-known fact of color
dynamics: Two color-{\bf 3} quarks have an attractive color-$\bar{\bf
3}$ channel that is fully half as strong from gluon exchange as the
attraction of a $\bar q (\bar{\bf 3}) q ({\bf 3})$ pair into a
color singlet.  This result follows from simple SU(3) color group
theory: The coupling of two representations $R_1$ and $R_2$ to a
representation $R$ is proportional to the combination of quadratic
Casimirs given by $C_2 (R) - C_2(R_1) - C_2 (R_2)$.  For two quarks,
the only attractive channels are the $\bar q q$ singlet ($R_1 = \bar
{\bf 3}$, $R_2 = {\bf 3}$, $R = {\bf 1}$) and the $q q$ anti-triplet
($R_1 = R_2 = {\bf 3}$, $R = \bar {\bf 3}$), and the latter is half as
strong as the former.  In this sense, diquarks are special entities in
QCD; if two quarks are created closer to each other than to any
antiquarks, then it is natural to expect them to form a bound
quasi-particle.  In fact, the greater the energy available in a system
(as often occurs in heavy-quark processes), the more opportunities
arise for such channels to occur.

As an explicit example, we recently proposed~\cite{Brodsky:2014xia} a
new paradigm: The tetraquarks are not simple quasi-static
$\delta$-$\bar \delta$ bound states, but instead arise as modes of a
rapidly separating $\delta$-$\bar \delta$ pair, remaining confined, a
color flux tube stretching between them with its length determined by
the available energy.  Many, but not all, of the narrow-width
tetraquarks lie near hadronic thresholds because these are the
energies at which the color string can easily break.  A case in point
is the $X(4632)$, the first exotic state found above the 4573~MeV
charmed-baryon ($\Lambda_c^+ \bar \Lambda_c^-$) threshold; it decays
dominantly to $\Lambda_c^+ \bar \Lambda_c^-$ (indeed, this is the only
mode yet seen), and this decay is precisely what one would expect from
flux-tube fragmentation to a single light $\bar q q$ pair.  Below this
threshold, the exotic state widths are not particularly large because
they can hadronize only by forming mesons with wave functions
stretching from the quarks in $\bar \delta$ to the antiquarks in the
$\delta$.  As suggested above, the $\delta$-$\bar \delta$ pair can
separate a significant distance ($> 1$~fm) if it is produced with
enough relative momentum; for instance, a $B$-meson decay can produce
this circumstance.

The decay of the $Z_c^+(4430)$ suggests that it is a spatially
extended state: Even though the $\psi(2S)$ and $J/\psi$ have precisely
the same $J^{PC} = 1^{--}$ quantum numbers, the $Z_c^+(4430)$ prefers
by a large margin to decay to $\psi (2S) \, \pi$ instead of $J/\psi \,
\pi$~\cite{Yuan:2014rta}; the $\psi(2S)$ mode has much less phase
space, but is spatially much larger than $J/\psi$, matching the
expected size from the diquark decay mechanism.

The diquark interpretation of tetraquarks naturally leads to the
possibility of more complex hadronic states in QCD, such as {\it
hexaquarks}~\cite{Bashkanov:2013cla}, which can arise as $\bar
\delta_{3C} \bar \delta_{3C} \bar \delta_{3C}$ color-singlets
analogous to $q_{\bar 3C} q_{\bar 3C} q_{\bar 3C}$ (anti)baryonic
bound states.  An example would be the charmed, charge $ Q=4$,
baryon-number $B=2$ state $[uu]_{\bar 3C} [cu]_{\bar 3C} [uu]_{\bar
3C} $.  In this case, two of the diquarks, {\it e.g.}, $[uu]_{\bar 3C}
$ and $ [cu]_{\bar 3C}$, can arrange themselves into a color-triplet
$(3C)$ four-quark cluster---precisely by the same analysis of
attractive color channels described above---which then, in turn, binds
to the $[uu]_{\bar 3C} $ diquark.  Thus, one can consider such
multiquark states as a {\em sequence of two-body bound-state
clusters\/} of color-triplet and anti-triplet states.  Another example
is the $B=2$ {\it octoquark\/} resonance $\left| \bar c c uud uud
\right>$, which can explain the dramatic spin dependence seen in
elastic $pp \to pp$ scattering at the charm production threshold
$\sqrt s \simeq 5$~GeV~\cite{Brodsky:1987xw}.  Again, this state can
be considered as sequential binding of four diquarks.  The light-front
wave function in this case satisfies a cluster decomposition analogous
to that in the structure of the deuteron~\cite{Brodsky:1985gs}.

In this paper, we put the $\delta$-$\bar \delta$ tetraquark picture to
a new dynamical test, using the well-known {\it constituent counting
rules\/}~\cite{Brodsky:1973kr,Matveev:1973ra,Brodsky:1974vy,
Farrar:1979aw,Efremov:1978rn,Duncan:1979hi,Duncan:1980qd,Lepage:1980fj,
Sivers:1982wk,Mueller:1981sg,Brodsky:1989pv}, to predict the
fixed-$\theta_{\rm cm}$ power-law scaling in Mandelstam $s$ of
exclusive processes at high energies.  The counting rules determine
the $s$ power-law dependence of cross sections and form factors for
processes at high $s$ and fixed scattering angle $\theta_{\rm cm}$,
where the power of $s$ is determined by the total number of
fundamental constituents---incoming plus outgoing---appearing in the
hard scattering.  The first application of the counting rules in this
tetraquark picture, a study of the so-called {\it cusp effect\/} of
threshold-induced shifts to resonance masses, appeared in a very
recent paper~\cite{Blitz:2015nra} by one of the present authors.

The counting rules, derived from the twist of the interpolating fields
controlling each hadron at short distances, provide two especially
interesting probes of tetraquark states.  First, as originally pointed
out in Refs.~\cite{Kawamura:2013iia,Kawamura:2013yya}, the high-$s$
production data for exotic states should follow the $s$ power
dependence predicted based on their expected valence quark structure
({\it i.e.}, four-quark tetraquarks).  Second, and original to this
work, the counting rules should be sensitive to the presence of
diquarks.  If particularly strongly-bound diquarks are formed in the
production process, and dissociate only after their production (so
that the diquarks effectively function for a time as single dynamical
units) then the counting rules will treat the diquarks effectively as
elementary fundamental color-triplet constituents at intermediate
energies.

We will also discuss a simple method of producing numerous exotic
states, via electroproduction near the charm (or other heavy-quark)
threshold.  While not directly dependent upon the diquark hypothesis,
the production mechanism also addresses the formation and dynamics of
QCD multiquark exotics.

This paper is organized as follows: In Sec.~\ref{sec:QuarkCount}, we
briefly review the constituent counting rules, how they are derived,
and their limitations.  Section~\ref{sec:Results} presents our
principal predictions for the production cross sections and the form
factors of exotic charmonium and bottomonium states.
Section~\ref{sec:Electro} proposes straightforward experimental
methods of producing numerous exotic states, particularly via
electroproduction processes.  In Sec.~\ref{sec:Concl} we summarize and
indicate future directions.

\section{Constituent Counting Rules} \label{sec:QuarkCount}

The constituent counting rules, which we briefly review in this
section, were developed in the decade subsequent to the creation of
perturbative QCD (pQCD)~\cite{Brodsky:1973kr,Matveev:1973ra,
 Brodsky:1974vy,Farrar:1979aw,Efremov:1978rn,Duncan:1979hi,
 Duncan:1980qd,Lepage:1980fj,Sivers:1982wk,Mueller:1981sg,
 Brodsky:1989pv}.  In essence, they represent the conformality and
scale invariance of QCD at high energies, and therefore are applicable
to a wide variety of field theories; for example, they have been
derived nonperturbatively in AdS/QCD~\cite{Polchinski:2001tt}.  The
summary presented here follows the more detailed introduction in
Ref.~\cite{Kawamura:2013iia}, which also was the first work to apply
counting rules to exotic multiquark hadrons.

The counting rules find their most incisive applications in
fixed-$\theta_{\rm cm}$ exclusive scattering processes at high $\sqrt
s$, for which none of the particles are accidentally close to being
collinear.  Constituent masses can then be neglected, and all of the
Mandelstam variables $s$, $t$, and $u \simeq -(s+t)$ are large.  To
maintain the integrity of the exclusive states, each of the
constituents must undergo a large momentum transfer to be deflected
through the same finite angle $\theta_{\rm cm}$, {\it i.e.}, fixed
$t/s$; therefore, all large energy scales may be expressed in terms of
$s$.  In pQCD, hard gluon exchanges are responsible for the momentum
transfers, while if leptons also appear in the process ({\it e.g.}, in
electroproduction), then hard electroweak gauge boson exchanges must
also be taken into account.  In the AdS/QCD picture, the ``operator
dictionary'' relates the counting rules to the short-distance twist
dimension of interpolating fields.

The counting rules in their simplest form simply enumerate $s$ factors
in propagators and spinor normalizations.  For simplicity, let us
begin with processes in which all $n$ external constituents, $n =
n_{\rm in} + n_{\rm out}$, are fermions.  In order for every
constituent to share an $O(1)$ fraction of the total $s$, the
leading-order Feynman diagrams for the scattering must have at least
$\frac n 2 - 1$ hard gauge boson propagators ($\sim 1/s$), and the
associated vertices much be connected by at least $\frac n 2 - 2$
internal constituent propagators ($\sim 1/\sqrt{s}$).  Since each
external constituent fermion field carries a spinor normalization
scaling as $s^{\frac 1 4}$, the fermion scaling factors cancel except
for an overall factor $s$, leaving a total invariant amplitude ${\cal
M}$ scaling as
\begin{equation} \label{eq:amplitude}
{\cal M} \propto 1/s^{\frac n 2 - 2} \, .
\end{equation}
The cross section for a scattering process in which the constituents
form two initial-state and two final-state particles may then be
written
\begin{equation} \label{eq:scaling}
\frac{d \sigma}{d t} = \frac{1}{16 \pi s^2} | {\cal M} |^2 \equiv
\frac{1}{s^{n - 2}} f \left( \frac{t}{s} \right) \, .
\end{equation}
As fixed by the mass dimension $M^{-4}$ of the left-hand side of this
equation, the function $f$ has mass dimension $M^{2n-8}$.  However,
$f$ itself is constructed so as not to scale with $s$; its
dimensionful factors instead arise from the amplitude overlaps between
the fundamental constituent fields and their external composite
states, such as those defining decay constants.

Modifying this result to allow for external gauge bosons is
straightforward: Each external boson line introduced replaces two
external fermion lines [$\sim (\sqrt{s})^2$] and one hard gauge boson
propagator ($\sim 1/s$), which cancel and therefore give precisely the
same scaling expressions for ${\cal M}$ and $d\sigma/dt$ as in
Eqs.~(\ref{eq:amplitude})-(\ref{eq:scaling}).

In the case of this work, we consider the scenario in which diquarks
are treated as fundamental constituents for purposes of counting,
since they are tightly bound to each other compared to their binding
with the other quarks; in scattering processes, they may be redirected
as a single unit.  Moreover, as discussed above, they form overall
color triplets in their channel of greatest attraction and therefore
can interact with the other quarks via single-gluon exchange.  If one
replaces the two dynamical quarks with a single diquark in the
counting argument, one loses the hard gluon connecting them ($\sim
1/s$), while two hard (fermionic) quark propagators are replaced by a
single (bosonic) diquark propagator [$(1/\sqrt{s})^2 \to 1/s$], and
the four external spinor normalizations [$(s^{\frac 1 4})^4$] are
replaced with two external diquark normalizations [$(\sqrt{s})^2$].
The latter two actions do not change the net counting of $s$ factors,
while the removal of the extra gluon leads to a scaling equivalent to
that of reducing the number of constituents from $n$ to $n-2$, exactly
as one would have from treating the diquark as a fundamental
constituent.  One must note, however, that this scaling holds only if
the diquark propagates intact through the scattering process, or if a
$\delta$-$\bar \delta$ pair is created in a pointlike configuration.

The original thrust of Ref.~\cite{Kawamura:2013iia} actually points to
a contrary but complementary direction: If an exotic multiquark state
is produced without a diquark component, then the scattering cross
section receives a contribution from all of the component valence
quarks.  The primary example discussed in~\cite{Kawamura:2013iia} is
$\pi^- + p \to K^0 + \Lambda(1405)$, where the small $\Lambda(1405)$
mass relative to non-strange analogues such as $N(1535)$ has led to
the proposal that the former is a pentaquark state.  In this case, one
can test this hypothesis by measuring whether $d\sigma/dt$ at large
$s$ scales according to Eq.~(\ref{eq:scaling}) as $s^{-(2+3+2+3-2)} =
s^{-8}$ or $s^{-(2+3+2+5-2)} = s^{-10}$.  The same authors
subsequently applied the large-$s$ scaling counting rules to examine
the properties of generalized parton distributions and distribution
amplitudes that appear in processes such as
these~\cite{Kawamura:2013yya}.  Whether diquarks act as one or two
fundamental constituents thus becomes an experimentally testable
prospect.

By applying Eq.~(\ref{eq:amplitude}), one can determine the large-$s$
behavior of hadronic form factors from the corresponding amplitude
${\cal M}$.  To give one explicit example, the $Z_c^+$ electromagnetic
form factor $F_{Z_c}(s)$ should scale as~\cite{Blitz:2015nra}
\begin{equation} \label{eq:formfactorscale}
F_{Z_c}(s) \to \frac{1}{s^{\frac 1 2 (1+1+4+4) - 2}} =
\frac{1}{s^3} \, ,
\end{equation}
but scale as $\sim 1/s^1$ if the $\delta$ and $\bar \delta$ are very
tightly bound.

A number of technical complications of real QCD modify the simple $s$
scaling naively obtained from perturbative Feynman diagrams.  Included
in this list are $\alpha_s (s)$ running and renormalization-group
effects in the parton distribution
amplitudes~\cite{Lepage:1979zb,Efremov:1979qk}, Sudakov
logarithms~\cite{Sudakov:1954sw,Cornwall:1975ty,Sen:1982bt}, ``pinch''
singularities from virtual gluons going on mass
shell~\cite{Landshoff:1974ew}, and singularities of the ``endpoint''
type when one or more constituents carry only a $\ll 1$ fraction of
the total $s$~\cite{Li:1992nu}.  Even so, the current consensus view
holds that the leading power-law scaling in $s$ for exclusive
processes remains the same as in the naive analysis.

\section{Scaling of Tetraquark Cross Sections} \label{sec:Results}

The scaling results and discussion of the previous section can be
directly applied to make a number of simple and experimentally
testable predictions for processes involving exotic states.  The most
straightforward applications use
Eqs.~(\ref{eq:amplitude})--(\ref{eq:formfactorscale}) to predict cross
sections at high-$s$.  Assuming that the $Z_c^+$ has four independent
fundamental constituents that share $O(1)$ fractions of the total
energy, one expects
\begin{equation}
  \frac{d\sigma}{dt} \left( e^+ e^- \to Z_c^+ (\bar c c \bar d u) +
    \bar Z_c^- (\bar c c \bar u d) \right) \propto \frac{1}{s^8} \, ,
\end{equation}
at finite scattering angle $\theta_{\rm c.m.}$.  The same scaling
occurs if four ordinary mesons are produced in a direct $e^+ e^-$
annihilation process.  On the other hand, if the $Z_c$ states are
formed from especially tightly-bound $\delta$ and $\bar \delta$
quasiparticles, the scaling exponent drops to $1/s^4$, which is the
same result as for the production of two ordinary mesons.  In that
case, it is then only because the exotic content of the $Z_c$ is well
established---the $Z_c$ clearly contains hidden charm, but is
nevertheless charged---that one can unambiguously assert the $Z_c$ is
not an ordinary meson.

We note at this point an interesting distinction if one selects the
production of neutral pairs $X \bar X$ or $Y \bar Y$, for which the
scaling arguments are the same as for $Z_c$ pairs.  If the  $X$ or $Y$
are comprised of tightly-bound diquarks, then one cannot be certain
from the scaling behavior alone whether or not the neutrals truly
carry exotic quark content.  Furthermore, by $Z_c$ we do not mean just
the $Z_c^+ (4430)$, although its possession of a dominant decay mode
$\psi(2S) \pi^+$ should make its reconstruction simpler.

Arguing against such a simple test is the fact that the scaling
predicted by the counting rules only holds when $s$ is ``large
enough.''  At very high values of $s$, one expects the scaling to work
well, but then one is faced with both a paucity of data and a
proliferation of final states from which to extract the exclusive
two-particle events.  At lower values of $s$, one is faced with the
problem that having more constituents requires a larger total $s$
before one is confident of all constituents carrying energies that lie
in the perturbative regime; just from counting alone, one would expect
the onset of scaling behavior for $e^+ e^- \to Z_c^+ \bar Z_c^-$ to
occur at an $s$ value about $(10/6)^2 \simeq 2.8$ times higher than
for $e^+ e^- \to$~meson~+~meson.  One can ameliorate this effect (not
to mention greatly increase the rate) by considering semi-exotic
processes such as $e^+ e^- \to Z_c^+ (\bar c c \bar d u) + \pi^- (
\bar u d)$, but the question of the precise onset point for the
asymptotic scaling regime remains.

A very simple modification, which extends the reach of the scaling to
lower $s$ is to form cross section ratios in order to eliminate
systematic corrections.  For example,
\begin{equation} \label{eq:Zc_ratio}
\frac{\sigma (e^+ e^- \to Z^+_c(\bar c c \bar d u) + \bar{Z}^-_{\bar
    c}(\bar c c \bar u d) )}{\sigma(e^+ e^- \to \mu^+ \mu^-)}
= \left| F_{Z_c}(s) \right|^2 \propto \frac{1}{s^{n-4}} \, ,
\end{equation}
where the exponent of $1/s$ is 6 if $Z_c$ is a bound state of a two
quarks and two antiquarks, and 2 if it is a bound state of two
particularly tightly-bound diquarks ($n$ referring only to the number
of constituents in the numerator process).  Indeed, the first equality
in Eq.~(\ref{eq:Zc_ratio}) is effectively a definition of the form
factor $F_{Z_c}(s)$, so it holds all the way down to the threshold, $s
= 4m_{Z_c}^2$.

In fact, one can perform a verification of the constituent content of
the $Z^+_c$ with a significantly larger rate by measuring a ratio that
gives the transition form factor
\begin{equation}
\frac{\sigma( e^+ e^- \to Z^+_c(\bar c c \bar d u) + \pi^-(\bar u d))}
{\sigma(e^+ e^- \to \mu^+ \mu^-)} = \left| F_{Z_c, \pi} (s) \right|^2
\propto \frac{1}{s^{n-4}} \, ,
\end{equation}
where the exponent of $1/s$ is 4 if $Z^+_c$ is a bound state of a two
quarks and two antiquarks, and 2 if it is a bound state of two
tightly-bound diquarks.

Another type of $e^+ e^-$ exclusive annihilation cross section ratio
is particularly interesting.  Consider the archetype process ratio:
\begin{equation} \label{eq:Z_Lambdac}
\frac{ \sigma (e^+ e^- \to Z^+_c(\bar c c \bar d u) + \pi^-(\bar u d)
)}
{\sigma (e^+ e^- \to \Lambda_c(cud) \bar{\Lambda}_c (\bar c \bar u
\bar d) ) } \, ,
\end{equation}
both of which have the same number of constituents, as well as the
same heavy-quark ($\bar c c$) constituents.  Therefore, both the
corrections due to high-$s$ scaling and corrections due to the total
heavy-quark mass cancel in this ratio.  While independent of $s$ at
leading order, whether the ratio is numerically large or small should
be sensitive to the fundamental QCD composition of the $Z_c^+ $ state;
to wit, if $Z_c^+$ is primarily a di-meson ``molecular''
hadro-charmonium $([\bar c c] + [\bar d u ] )$ or $\bar D$-$D$ $([\bar
c u] [\bar d c])$ state, it should be bound by weaker color-singlet
van der Waals forces, and thus be numerically smaller than if the four
quarks in the $Z_c^+$ remain coupled through color-nonsinglet hard
gluon exchanges.  Indeed, one can envisage truly peculiar scenarios:
If $Z_c^+$ contains tightly-bound diquarks but $\Lambda_c$ does not,
then the ratio of Eq.~(\ref{eq:Z_Lambdac}) could actually grow with
$s$ (in this case, as $s^2$).

Finally, we have to this point considered only $e^+ e^-$ collider
processes.  Similar final states produced through $\bar p p$
annihilation have identical ratios as powers of $s$.  The absolute
high-energy cross sections fall with an additional power of $s^4$
(three quarks in each hadron, as compared with a lepton-antilepton
pair), but the rates can also be greatly enhanced due to $\bar p p$
already containing the light quarks required by the final state.  To
be explicit,
\begin{equation}
\frac{d\sigma}{dt} \left( \bar p ( \bar u \bar u \bar d) p ( u u d)
\to Z_c^+ ( \bar c c \bar d u) + \pi^- ( \bar u d) \right) \propto
\frac{1}{s^{10}} \, ,
\end{equation} 
should be numerically substantial near threshold and fall off very
quickly for large $s$, while the ratio
\begin{equation}
\frac{\sigma(\bar p p \to Z_c^+ \pi^-)}
{\sigma(\bar p p \to \Lambda_c \bar \Lambda_c)} \,
\end{equation}
has again the same quark content in numerator and denominator, and
therefore again has cancelling scaling and heavy-quark content
factors; in particular, its dependence on $s$ should be much gentler.

Of course, many of the charmed processes analogous to those described
here and below have direct analogues in the $\bar b b$ threshold
region; for a discussion from a different theoretical perspective, see
Refs.~\cite{Karliner:2008rc,Karliner:2011yb,Karliner:2014lta}.

\section{Electroproduction of exotic states near threshold}
\label{sec:Electro}

The existence of the first genuine QCD exotics having apparently been
experimentally confirmed, one is led to ask what other exotics await
discovery and what processes can most effectively be used to produce
them.  Here we argue that electroproduction near the charm threshold
provides a natural laboratory for creating such states.  This
kinematical region represents an obvious energy regime for the
formation of exotic states, because the slowly-movi
ng $\bar c$ and $c$
quarks produced readily coalesce with comoving valence quarks of the
target.  The diquark hypothesis is not essential to the analysis
presented in this section.

For example, consider the process $e p \to e^\prime X$ in the
target-proton rest frame, which is most naturally considered a
$\gamma^* p$ collision.  The virtual photon produces a $\bar c c$ pair
a significant fraction of the time.  A simple estimate for the ratio
of $\bar c c$ to $\bar u u$ pair-production events is the ratio of
$s$-channel squared masses at threshold, which is $ \sim \frac{(m_p+
  m_\pi)^2}{(m_{\Lambda_c}+ m_D)^2} \simeq 7\%$ at fixed $\gamma^*$
momentum transfer above the charm threshold; note that using the
slightly lower hidden-charm threshold $(m_p + m_{J/\Psi})^2$ in the
denominator produces almost the same result.  For energies slightly
above the charm-production threshold, $s = (q + p)^2 \simeq
(4.2$~GeV)$^2$ where $q$ and $p$ are, respectively, the photon and
proton momenta, then the charm quarks coalesce with the $uud$ valence
quarks moving at the same rapidity to produce open-charm states such
as $\gamma^* p \to \overline D^0 (\bar c u) \Lambda^+_c(cud)$.  For $s
>$~(4.0~GeV)$^2$, the hidden-charm process $\gamma^* p \to J/\Psi + p$
remains possible.  Analysis such as decribed in
Refs.~\cite{Anikin:2003fr,Anikin:2005ur} for $\gamma^* \gamma \to \rho
\rho$ may then be peformed.

But electroproduction also provides a direct way to produce exotic
hadronic states such as the $Z^+_c (\bar c c \bar d u) $ tetraquark
and the ${\cal O}(\bar c c uud uud)$ octoquark.  For example, if a
low-mass {\it pentaquark\/} ${\cal P}(\bar c cuud)$ exists below the
$\overline D \Lambda_c$ threshold, then the process $e p \to e^\prime
{\cal P}(\bar c cuud)$ would occur, with ${\cal P}$ manifesting as a
peak in the missing-mass ($M_X$) distribution of $e p \to e^\prime X$.
If $4.2 \ {\rm GeV} > m_{\cal P} > 4.0 \ {\rm GeV}$, then ${\cal P}$
would likely appear as a resonance decaying to $J/\psi + p$.  And if
$m_{\cal P} < 4.0$~GeV, then ${\cal P}$ would exist as a bound state
and appear as a sharp peak in the $M_X$ distribution of $e p \to
e^\prime X$.

Alternately, one can perform an indirect search for a ${\cal P}$ thus
produced; if ${\cal P}$ has a sufficiently long lifetime, it can
collide with a second nucleon in a fixed target downstream from the
initial collision point and materialize as a hidden- or open-charm
state,
\begin{eqnarray}
  e + p & \to & e^\prime + {\cal P} \, , \nonumber \\ & &
  \hspace{2.0em} \Downarrow \nonumber \\ & & \hspace{1.9em}
  {\cal P} + N \to N J/\psi , \ \overline D \Lambda_c \, .
\end{eqnarray}

Even more exotic states could be produced this way, if they indeed
exist.  A very interesting experimental signal dating back three
decades~\cite{Court:1986dh} is the surprisingly large spin-spin
correlation in $pp$ elastic scattering, sometimes called the {\it
  Krisch effect}.  The polarized cross sections for scattering of
protons with spins normal to the scattering plane have a remarkable
asymmetry: At $s = ( 5 \ {\rm GeV})^2$,
\begin{equation}
  \frac{\frac{d\sigma}{dt} ( p_\uparrow p_\uparrow \to p p )}
  {\frac{d\sigma}{dt} ( p_\uparrow p_\downarrow \to p p )} \simeq 4 \,
  .
\end{equation}
Such an asymmetry is strongly at odds with the expectations of pQCD,
since at such high energies one expects spin differences to be washed
out.  Note, however, that such an effect can occur if the high-energy
process interferes with a resonance lying right at $s = ( 5 \ {\rm
  GeV})^2$.  Since the baryon-number $B = 2$ hidden-charm threshold is
$\simeq 2m_p + m_{J/\Psi} = 5.0$~GeV, the production of an {\it
  octoquark} state ${\cal O}^{++} = \left| \bar c c uud uud \right>$
has been proposed as a resolution~\cite{Broadsky:2012rw}.  Should such
a state exist, one can use the electroproduction methods to search for
its isospin partner ${\cal O}^+$ in the missing-mass spectrum of
collisions on a deuteron $d$ target:
\begin{equation}
e + d \to e^\prime + {\cal O}^+ (\bar c c uud udd) \, ,
\end{equation}
which would appear in the missing-mass spectrum of $e d \to e^\prime
X$.  In this case, the open- and hidden-charm thresholds lie at $M_X
\simeq m_{\Lambda_c} + m_n + m_{D^0} = 5.1$~GeV and $m_p + m_n +
m_{J/\Psi} = 5.0$~GeV, respectively, and comments analogous to the
ones above for the ${\cal P}$, regarding whether the ${\cal O}^+$
would appear as a resonance or a bound state, apply here as well.
Furthermore, if ${\cal O}^+$ is long-lived, it could decay to $J/\psi
+ p + n$ or could be dissociated as ${\cal O}^+ \! + A \to J/\psi + p
+ n + A^\prime$ in subsequent collisions in a nuclear target.

These methods for finding exotic states in electroproduction can be
extended to the production of {\it nuclear-bound
quarkonium}~\cite{Brodsky:1989jd,Luke:1992tm} states $[\bar c c] A$,
in which quarkonium is bound to nuclei by QCD van der Waals
interactions---the nuclear analogue to hadro-charmonium.  Such states
could be produced in $e A \to e^\prime X$ collisions.

What of the original tetraquarks?  If the outgoing baryon in the
electroproduction process $e p \to e^\prime X$ is a $p$, then the
extra inelastically-produced state is neutral.  Since the $X(3872)$ is
the best-characterized exotic state, perhaps a natural place to start
the electroproduction program is by observing the process $e p \to
e^\prime p^\prime X(3872)$ near its $s = (4.8$~GeV)$^2$ threshold.  As
for the charged tetraquarks such as $Z_c^+(4430)$, a charge-exchange
electroproduction process $e p \to e^\prime n Z_c^+(4430)$ is required,
which at its core can be considered a $\gamma^* \pi^+$ collision.
Neutral ${\cal P}$ and ${\cal O}$ states could be created analogously,
at correspondingly higher thresholds.

To date, all of the observed exotic candidates contain either hidden
charm or bottom.  Is incorporating heavy quarks a necessary feature of
observable tetraquark states?  The original $\delta$-$\bar \delta$
mechanism presented in Ref.~\cite{Brodsky:2014xia} depends upon having
sufficient energy release in the production process that the $\delta$
and $\bar \delta$ separate far enough so as to be considered distinct
particles; perhaps for lighter systems the diquarks try to form but
dissolve immediately into meson pairs.  In any case, all of the
searches described above apply to the strange sector as well, such as
a $u u d \bar s s$ pentaquark.  One possible result is that
hidden-charm exotics emerge naturally from these electroproduction
experiments but hidden-strangeness ones do not.

As an intermediate case, one can also study open-charm,
open-strangeness states using $e p \to e^\prime \Lambda X$ as a
$\gamma^* K^+(u \bar s) $ collider.  In this case, one would produce
charged charmed-strange tetraquarks such as $\bar c c \bar s u$.
Here, one would look for peaks in the $M_X$ distribution after tagging
the final-state electron $e^\prime$ and $\Lambda$ baryon.

Another interesting case is $e^+ e^-$ annihilation to four heavy
quarks.  For example, just above the $\bar c c \bar c c$ threshold,
one can produce $e^+ e^- \to J/\psi ~\eta_c$.  Just below threshold,
the four heavy quarks can rearrange to form a exclusively charmed
tetraquark as a bound state of $[cc]_{\bar 3_C}$ and $[\bar c\bar
c]_{3_C}$ diquarks.

\vspace{3ex}

\section{Discussion and Conclusions} \label{sec:Concl}

We have proposed a number of experimentally straightforward and
feasible tests of the exotic nature of the recently-discovered
tetraquark candidates such as $X(3872)$ and $Z_c^+(4430)$.  Scenarios
in which the four quarks independently carry ${\cal O}(1)$ fractions
of the hadron momentum, and scenarios in which the four quarks are
segregated into tightly-bound diquark and antidiquark pairs, have been
explored utilizing constituent counting rules, which are normally
limited to tests at high momentum transfer.  By forming ratios of
cross sections to different exclusive states, one can extend the
usefulness of the counting rules to the threshold domain for producing
heavy exotic states.

We have also discussed several promising methods to produce other
exotic multiquark states in near-threshold electroproduction and
electron-positron annihilation.  One can also confirm the existence of
known exotic $\bar c c$ states by creating them just above the
threshold for production of the charm-quark pair, where the limited
phase space makes the formation of exotics likely, through coalescing
the soft heavy quarks with the light valence quarks moving at similar
rapidities.

The exotic hadron production processes discussed in this paper lead to
many new experimental opportunities at $e^+ e^-$ colliders such as BES
and Belle, at electroproduction facilities such as the 12~GeV Upgrade
of JLab and proposed $ep$ colliders, and at new hadronic beam
facilities such as $\bar{\rm P}$ANDA at FAIR and AFTER@LHC.

QCD, now in its fifth decade, continues to present us with surprises.
Even the full extent of its basic hadronic spectrum remains an open
question.  However, given the results of sufficiently ingenious
experiments, an ever deepening understanding of the theory and its
novel features will inevitably follow.


\begin{acknowledgments}
  This work was supported by the U.S.\ Department of Energy under
  Grant No.\ DE-AC02-76SF00515 (S.J.B.) and by the National Science
  Foundation under Grant Nos.\ PHY-1068286 and PHY-1403891 (R.F.L.).
  In addition, S.J.B.\ thanks V.~Ziegler for discussions on the
  feasibility of experimental measurements of the processes discussed
  here and S.~Kumano for discussions of his work.
\end{acknowledgments}


\end{document}